\documentclass{article} \usepackage{epsf} \usepackage{amsmath}
\usepackage{amssymb}
 
\newcommand{\be}{\begin{equation}} 
\newcommand{\ee}{\end{equation}}

\newcommand{\half}{\frac{1}{2}}

\newcommand{\bsplit}{\begin{equation} \begin{split}}
    \newcommand{\esplit}{\end{split} \end{equation}}

\begin{document}
\thispagestyle{empty} \parskip=12pt \raggedbottom
 
\def\mytoday#1{{ } \ifcase\month \or January\or February\or March\or
  April\or May\or June\or July\or August\or September\or October\or
  November\or December\fi
\space\number\day , \\
\space \number\time , \\
  \space \number\year}
\noindent
\hspace*{6cm}
\vspace*{1cm}
\begin{center}
  {\LARGE The QCD rotator in the chiral limit}
 
  \vspace{1cm} P. Hasenfratz\\
  Institute for Theoretical Physics \\
  University of Bern \\
  Sidlerstrasse 5, CH-3012 Bern, Switzerland
  
  \vspace{0.5cm}
  
  \nopagebreak[4]
 
\begin{abstract}
The low lying spectrum of QCD in the $\delta$-regime is calculated here in
chiral perturbation theory up to NNL order.The spectrum has a simple form in
terms of the pion decay constant $F$ and a combination of the low energy
constants $\Lambda_1$ and $\Lambda_2$. 
Since  measuring low
lying stable masses is among the easiest numerical tasks, the results 
should help fixing these parameters to good precision.
\end{abstract}
 
\end{center}
\eject

\section{Introduction and summary}
The low lying spectrum of QCD, in a special environment, is that of a simple
quantum mechanical rotator~\cite{Le1}. 
Although amazing as it is, this
spectrum will never be measured in QCD experiments. This
physics, however can be studied numerically which will deliver
precise predictions on some of the low energy constants in chiral perturbation
theory~\cite{SW,JGHL1,JGHL2}. 
One of the reasons is that measuring low
lying stable masses is among the easiest numerical tasks. Further, the QCD rotator
lives in a box of size $L_s \times L_s \times L_s \times (L_t \rightarrow
\infty)$, which creates an infrared-safe environment\footnote{It is known 
since a long time 
in condensed matter physics also that in a finite box at zero temperature the
lowest excitations in the spontaneously broken phase are related to the slow 
precession of the order parameter described by a rotator
ref.~\cite{FiPr,BEZJ}.}. This allows to study the
chiral limit first and switching on the symmetry breaking terms later. In
two-flavor QCD, in the leading order ($L$) of chiral perturbation theory the
$SU(2) \times SU(2) \sim O(4)$ rotator has an inertia $\Theta$ proportional to 
the size of the spatial box: $\Theta=F^2 L^3_s$. Here $F$ is the pion decay
constant in the chiral limit. In the next-to-leading order ($NL$) the inertia 
is corrected $\Theta=F^2 L^3_s(1+\sim 1/F^2L^2_s)$~\cite{PHFN}, 
where $1/F^2 L_s^2$ is the small expansion parameter. 
The logarithms and the additional low
energy constants $l_1$ and $l_2$ enter only in the $NNL$ order, which is the
highest order we consider in this work.   

In the environment discussed above ($\delta$-regime~\cite{Le1}) it is natural
to divide 
the degrees of freedom into $fast$ and $slow$ modes. The fast modes can be
treated in perturbation theory, while the slow modes build the slowly moving
rotator, whose energy excitations are much smaller than those of the standard
Goldstone boson excitations which carry finite momenta.

Let us summarize the final results before going over the details.
We quote the result for
$N=4$, which corresponds to 2-flavor QCD.
Up to $NNL$ order the rotator spectrum in the chiral limit has the form
\footnote{Terms with other Casimir forms are expected to enter beyond NNL
  order.} 
\begin{equation}
\label{1}
E_l=\frac{1}{2\Theta} l(l+2) \,,\,\, l=0,1,2,\dots \,,
\end{equation}
where $N$ refers to the underlying group $O(N)$. The corrections from the 
perturbative expansion enter in the
inertia $\Theta$ which, up to $NNL$ order in the chiral limit, reads
\be
\label{2}
\begin{split}
\Theta  = F^2 L^3_s & \left\{  1-\frac{2}{F^2 L^2_s}{\bar G}^{*} \right. \\
& + \frac{1}{(F^2 L^2_s)^2} \left[ 0.088431628 \phantom{\frac14} \right. \\
& + \left. d0d0{\bar G}^{*}\frac{1}{3 \pi^2} 
\left. \left( \frac14 \ln(\Lambda_1L_s)^2 + 
\ln(\Lambda_2L_s)^2 \right)  
\right] \right\} \,.
\end{split}
\ee

Here $\Lambda_1,\Lambda_2$ are the standard scales
related to the bare low energy constants $l_1,l_2$ in the chiral
Lagrangian. The conventions for $\Lambda_1,\Lambda_2$ are 
given in section~7.

We are using dimensional regularization (DR) in this work.
The constants ${\bar G}^{*}$ and $d0d0{\bar G}^{*}$ are related to the
constrained Green's function ${\bar D}^{*}(x)$ and its second time derivative
$\partial_0 \partial_0 {\bar D}^{*}(x)$, which enter the perturbation theory:
\begin{equation}
\label{3}
{\bar D}^{*}(0)  =\frac{1}{L^2_s}{\bar G}^{*} \,,\,\,\, 
{\bar G}^{*}=-0.2257849591 \,.
\end{equation}

\begin{equation}
\label{4}
\partial_0 \partial_0 {\bar D}^{*}(0) =\frac{1}{L^4_s}d0d0{\bar G}^{*}\,,\,\,\,
d0d0{\bar G}^{*}=-0.8375369106 \,\,.
\end{equation}

In the Green's function ${\bar D}^{*}(x)$ the non-perturbative slow modes 
(rotator modes) are missing (notation $D^*$) and the UV singularity is also
subtracted (notation $\bar{D}$). The precision of the
numerical numbers are estimated to be $10^{-9}$, or better\footnote{Of course,
  no such precision is needed in this perturbation theory. It might help
  though when comparing with related works in the future.}. Definitions and
properties of the Green's functions we use are summarized in section 5.

Although the final result in eq.~\eqref{2} is very simple, the underlying 
chiral perturbation
theory is not. For this reason it is a good news that an independent
calculation is under way using a completely different
technique~\cite{FNCW}. Including the symmetry breaking contributions is also
in progress~\cite{MW}.

\section{The chiral action and the rotator in leading order}
The low energy limit of QCD with two massless quarks $m_u=m_d=0$ is described
by an effective non-linear $O(N=4)$ $\sigma$-model. The Lagrangian, up to
1-loop level, has the form $L_\mathrm{eff}= L^{(2)}_\mathrm{eff} +
L^{(4)}_\mathrm{eff}$, 
where\footnote{We assume
that rotation symmetry is respected by the regularization. In
dimensional regularization this is the case.}
\be
\label{5}
\begin{split}
L^{(2)}_\mathrm{eff} & =\frac{F^2}{2} \partial_\mu \mathbf{S}\, 
\partial_\mu \mathbf{S} \,\,, \\
L^{(4)}_\mathrm{eff} &=-l_1\,(\partial_\mu  \mathbf{S}\, \partial_\mu
\mathbf{S})(\partial_\nu \mathbf{S}\, 
\partial_\nu \mathbf{S}) -l_2\,(\partial_\mu \mathbf{S} \,\partial_\nu
\mathbf{S})(\partial_\mu \mathbf{S}\, 
\partial_\nu \mathbf{S})\,.
\end{split}
\ee
Here $F,l_1,l_2$ are the bare low energy constants and the $N$-component field has
unit length 
\begin{equation}
\label{6}
S_a(x), \,\, a=0,1,\dots,(N-1), \,\,\, \mathbf{S}^{2}(x)=1 \,.
\end{equation}

The action in eq.~\eqref{5} might also be interpreted as the effective low
energy prescription of a ferromagnet. We consider an $L_s \times L_s \times
L_s \times L_t$ box with spatial volume $V_s=L^3_s$, while the (Euclidean) time
extension is taken very large, $L_t \rightarrow \infty$. In the $L_s
\rightarrow \infty$ limit the system has a net magnetization and massless
Goldstone bosons.

We consider a cylinder geometry ($\delta$-regime), where $L_s$ is finite,
but sufficiently large so that the (would be) Goldstone bosons dominate the
finite size effects. Due to the microscopic magnetic moments, the $L_s \times
L_s \times L_s$ spatial box has a net magnetization on each time slices. Since
$L_s$ is finite, this net magnetization is moving around as 
the function of the time $t$.

In leading order $(L)$, which is a good approximation if the dimensionless
expansion parameter $1/(F^2 L^2_s)$ is small, the microscopic
magnets on a time slice are parallel and the rotator action 
from eq.~\eqref{5} goes over to
\begin{equation}
\label{7}
A_\mathrm{rot}=\frac{F^2 V_s}{2} \int dt
\,\dot{\mathbf{e}}(t)\dot{\mathbf{e}}(t) \,. 
\end{equation}  
Here ${\mathbf{e}}(t)$ is the direction of the total magnetization in the
internal $O(N)$ space at the time $t$:
\begin{equation}
\label{8}
{\mathbf{e}}(t)=[e(t)_0,e(t)_1,\dots,
 e(t)_{N-1}],\,\,\,{\mathbf{e}}(t)^2=1 \,.   
\end{equation}  
Equation~\eqref{7} describes a quantum mechanical $O(N)$ rotator, the QCD
rotator in leading order~\cite{Le1},
with a discrete energy spectrum above the ground state:
\begin{equation}
\label{9}
\mathbf{H}=\frac{\mathbf{L^2}}{2\Theta} \,\,,\,\, \Theta=F^2 V_s \,,\,\,
E_l=\frac{1}{2\Theta}\,l(l+N-2)\,,\,\,\, l=0\,1\,\,,2\,\dots \,. 
\end{equation}  
In eq~\eqref{9} $\mathbf{H}$ is the Hamiltonian, $\mathbf{L}$ is 
the $O(N)$ angular momentum and
$\Theta$ is the inertia of the rotator. Our aim is to determine the corrections
up to $NNL$ order, where the low energy constants $l_1,\,\,l_2$ first enter.

\section{Separating the slow and fast modes}

\subsection{New integration variables}
If $L_t \sim L_s$, it is sufficient to take special care of the freely
rotating total magnetization~\cite{JGHL3,JGHL4,JGHL5,PHHL}. 
For $L_t/L_s$ large, however the
magnetization on distant time slices in the cylinder might differ
significantly. We have to treat these special slow modes
non-perturbatively. These modes are the $k=(k_0,\mathbf{k}=\mathbf{0})$ 
modes in Fourier space. In this subsection we deal mainly with the measure, 
while the action is treated in Sections 3.2 and 3.4. 

The steps followed here are similar to that used in~\cite{PHFN}, but it is simpler
and exact in every order. Unlike in~\cite{PHFN}, where lattice regularization
was used, we apply here dimensional regularization.

Insert into the path integral the identity
\begin{equation}
\label{10}
1=\prod_t \int d{\mathbf{m}}(t) \prod_{a=0}^{N-1}
\delta[m^a(t)-\frac{1}{V_s}\int_{\mathbf{x}} S^a(t,\mathbf{x})]\,\,,
\end{equation}
where $\mathbf{x}$ is the spatial coordinate, $x=(t,\mathbf{x})$, and
\begin{equation}
\label{11}
{\mathbf{m}}(t)=m(t)\,{\mathbf{e}}(t)\,, \qquad{\mathbf{e}}^2=1\,,\qquad
d{\mathbf{m}}(t)=m(t)^{N-1} dm(t)\,d{\mathbf{e}}(t)\,\,. 
\end{equation}
The $O(N)$ vector of unit length ${\mathbf{e}(t)}$ is the direction of the
'magnetization' $\int_{\mathbf{x}}{\mathbf{S}}(t,\mathbf{x})$ on the time slice
$t$. The local 'magnets' ${\mathbf{S}}(t,\mathbf{x})$ fluctuate around the
slow mode ${\mathbf{e}}(t)$. 

Introduce the $O(N)$ rotation $\Omega$(t) as
\begin{equation}
\label{12}
{\mathbf{e}}(t)=\Omega(t) {\mathbf{n}}\,,\,\,\,\,\,
{\mathbf{n}}=(1,0,\dots,0)\,\,. 
\end{equation}
Having an $O(N)$ matrix $\Omega$ rather than a vector will be convenient in the
manipulations below. We consider $\Omega$ as a function of 
${\mathbf{e}}(t)$. Note, however, that the path integral
\begin{equation}
\label{13}
\begin{split}
Z = & \prod_x \int d{\mathbf{S}}(x)\, \delta({\mathbf{S}^2}(x)-1)\prod_t \int
dm(t)\,m(t)^{N-1} \int d{\mathbf{e}}(t) \\
    &\delta^{N}[m(t)\Omega(t){\mathbf{n}}-
        \frac{1}{V_s}\int_{\mathbf{x}}{\mathbf{S}}(t,\mathbf{x})]\,
e^{-A_\mathrm{eff}({\mathbf{S}})} \,,
\end{split}
\end{equation}
depends only on the first column of the matrix $\Omega$.

Introduce new integration variables ${\mathbf{R}}$ in the path integral:
\begin{equation}
\label{14}
{\mathbf{S}}(t,\mathbf{x})=
\Omega(t){\Sigma}(t)^T{\mathbf{R}}(t,\mathbf{x})\,\,, 
\end{equation}
where ${\Sigma}(t)\in O(N)$ and ${\mathbf{R}}$ has unit length.
The matrix ${\Sigma}(t)$ is taken to have a special structure:
${\Sigma}(t)_{ij}= \bar{\Sigma}(t)_{ij}$, 
${\Sigma}(t)_{00}=1$, $\bar{\Sigma}(t)_{0i}=0$, $\bar{\Sigma}(t)_{i0}=0$, where
$\bar{\Sigma}(t)$ is an $O(N-1)$ matrix. This $O(N-1)$ matrix
$\bar{\Sigma}(t)$ 
will be chosen conveniently later.

The partition function reads
\be
\label{15}
\begin{split}
Z = & \prod_x \int d{\mathbf{R}}(x)\, \delta({\mathbf{R}^2}(x)-1)\prod_t \int
dm(t)\,m(t)^{N-1} \int d{\mathbf{e}}(t) \\
    &\delta^{N}[m(t)\Omega(t){\mathbf{n}}-\Omega(t)
    {\Sigma}(t)^{T}\frac{1}{V_s}\int_{\mathbf{x}}{\mathbf{R}}
            (t,\mathbf{x})]\,e^{-A_\mathrm{eff}(\Omega{\Sigma}^T{\mathbf{R}})}
            \,.   
\end{split}
\ee
Using $\delta^{N}(\Omega{\mathbf{z}})=\delta^{N}({\mathbf{z}})$ and 
${\Sigma} {\mathbf{n}}={\mathbf{n}}$ gives
\be
\label{16}
\begin{split}
Z = & \prod_x \int d{\mathbf{R}}(x) \delta({\mathbf{R}^2}(x)-1)\prod_t \int
dm(t)m(t)^{N-1} \int d{\mathbf{e}}(t) \\
    &\delta^{N}[m(t){\mathbf{n}}-
    \frac{1}{V_s}\int_{\mathbf{x}}{\mathbf{R}}
            (t,\mathbf{x})]\,e^{-A_\mathrm{eff}(\Omega{\Sigma}^T{\mathbf{R}})}
            \,.   
\end{split}
\ee
eq.~\eqref{16} shows that on each time slice the vector ${\mathbf{R}}$ 
fluctuates around the $(1,0,\dots,0)$ internal direction:
\begin{equation}
\label{17}
{\mathbf{R}}(t,\mathbf{x})= (1,0,\dots,0) + {\rm small \,\,fluctuations}\,.
\end{equation}
The small fluctuations can be treated in perturbation theory:
\begin{equation}
\label{171}
{\mathbf{R}}(t,\mathbf{x})= [\,(1-{\mathbf{\Pi}^2}(t,\mathbf{x}))^{\frac{1}{2}}
\,,{\mathbf{\Pi}}(t,\mathbf{x})\,]\,,  
\end{equation}
where
\begin{equation}
\label{172}
 {\mathbf{\Pi}}(t,\mathbf{x})
=[\,\Pi(t,\mathbf{x})_1,\dots,
 {\Pi}(t,\mathbf{x})_{N-1}\,]
\end{equation}
are the small fluctuations.
From eqs.~\eqref{16},~\eqref{171} follows
\begin{equation}
\label{173}
\frac{1}{V_s} \int_{\mathbf{x}} \Pi(t,\mathbf{x})_i = 0,\qquad i=1,\dots,N-1\,\,.
\end{equation}

The field ${\mathbf{\Pi}}(x)$ is small. This fact allows to integrate out the 
${\mathbf{\Pi}}$-fields ('fast modes') in a systematic perturbation theory. We
should keep in mind that the slow modes ${\mathbf{e}}(t)$ were separated. The 
$k=(k_0,\mathbf{k}=\mathbf{0})$ modes in Fourier-space are slow and are not
part of the fast $\Pi$-fields.

Two of the intergrals in eq.~\eqref{16} are zero in dimensional
regularization. The first one is the well known measure which is created as we
replace  ${\mathbf{R}}$ by the ${\mathbf{\Pi}}$ integration variable. 
The second integral, when put in the exponent, has the form 
$(N-1)\int_t
\ln[1/V_s\int_{\mathbf{x}}((1-{\mathbf{\Pi}^2}(t,\mathbf{x}))^{1/2})]$,
which is zero also. The relevant part of the partition function $Z$ will be given
in Section 4. 

\subsection{The slow and fast modes in the action 
{$A^{(2)}_\mathrm{eff}(\Omega {\Sigma}^T {\mathbf{R}})$}}
Here and in Sec. 3.4 we consider the $\it action$ which is a classical
object. In the following steps we found it useful to introduce an infinitesimal
parameter $\epsilon$ which disappears at the end.

In Sec. 3.1 we introduced new variables
${\mathbf{S}}(t,\mathbf{x})=\Omega{\Sigma}^T{\mathbf{R}}(t,\mathbf{x})$. Let us
write the time derivatives in the leading action $A^{(2)}_\mathrm{eff}$ in the form
\begin{equation}
\label{18}
 \partial_t {\mathbf{S}}(t,{\mathbf{x}})\partial_t {\mathbf{S}}(t,{\mathbf{x}})=
 \frac{2}{\epsilon^2}[1-{\mathbf{S}}(t+\epsilon,\,{\mathbf{x}})
{\mathbf{S}}(t,{\mathbf{x}})]_{\epsilon \rightarrow 0}.
\end{equation}
 Eqs.~\eqref{14},~\eqref{18} imply
\begin{equation}
\label{19}
\partial_\mu \mathbf{S}(t,\mathbf{x})\partial_\mu \mathbf{S}(t,\mathbf{x})=
\partial_\mu \mathbf{R}(t,\mathbf{x})\partial_\mu \mathbf{R}(t,\mathbf{x}) -
\frac{2}{\epsilon^2} Q(t)\mathbf{R}(t+\epsilon,\mathbf{x})\,
\mathbf{R}(t,\mathbf{x})_{\epsilon \rightarrow 0}\,\,,
\end{equation}
where the $N \times N$ matrix $Q(t)$ is expressed in terms of $\Omega$ and
${\Sigma}$ 
\begin{equation}
\label{20}
Q(t)=V(t+\epsilon){\Sigma}(t+\epsilon)^T - 1\,.
\end{equation}
Here we introduced the notation
\begin{equation}
\label{21}
V(t+\epsilon)= {\Sigma}(t)\Omega(t)^T \Omega(t+\epsilon)\,\,.
\end{equation}
The $O(N)$ matrix $\Sigma(t)$ is defined in terms of the $O(N-1)$ matrix 
$\bar{\Sigma}(t)$ (see after eq.~\eqref{14}). We fix this matrix now:
\begin{equation}
\label{22}
\bar{\Sigma}(t)_{ij} = V(t)_{ij} - \frac{V(t)_{i0}V(t)_{j0}}{1+V(t)_{00}}\,.
\end{equation}
It is easy to show that $\bar{\Sigma} \in O(N-1)$, indeed. Further, $Q(t)$ can
be expressed now in terms of $V(t+\epsilon)$:
\be
\label{23}
\begin{split}
& Q(t)_{00}= V(t+\epsilon)_{00}-1 \,,\,\,
Q(t)_{0i}=-V(t+\epsilon)_{i0}\,,\,\,Q(t)_{i0}=V(t+\epsilon)_{i0}\,,\\
& Q(t)_{ij}=  -\frac{V(t+\epsilon)_{i0}
V(t+\epsilon)_{j,0}}{1+V(t+\epsilon)_{00}}\,.
\end{split}
\ee

Assume now that $V$ is known at some $t_0$, while $\Omega(t)$ is known for 
any $t$\footnote{It will soon turn out that only the first column (i.e. the
vector ${\mathbf{e}})$ is needed.}. Having $V(t_0)$, we can fix
${\bar{\Sigma}}(t_0)$ and  $\Sigma(t_0)$ using eq.~\eqref{22}.
Actually, we have the following chain
\begin{equation}
\label{24}
V(t_0) \underset{eq.~\eqref{22}}\longrightarrow\, \Sigma(t_0)
\underset{eq.~\eqref{21}}\longrightarrow\, V(t_0+\epsilon)
\underset{eq.~\eqref{22}}\longrightarrow\, \Sigma(t_0+\epsilon)
\underset{eq.~\eqref{23}}\longrightarrow\, Q(t_0)\,.
\end{equation}
At the and of this chain we have $Q(t_0)$. In addition, we have
$V(t_0+\epsilon)$, so we start the chain again creating $Q(t_0+\epsilon)$ and
so on.

We shall see that for the rotator spectrum up to NNL order the following three
combinations of the matrix $Q$ are needed only:
$Q(t)_{00}\,\,,Q(t)_{i0}Q(t)_{i0}$ and $Q(t)_{i,i}$ (the repeated index $i$
is summed). These combinations can be expressed in terms of the slow modes 
${\mathbf{e}}(t)$ in the ${\epsilon \rightarrow 0}$ limit: 
\begin{equation}
\label{25}
\begin{split}
& \frac{1}{\epsilon^2}Q(t)_{00}=
  -\frac{1}{2}\dot{\mathbf{e}}(t)\dot{\mathbf{e}}(t)\,,\,\,
\frac{1}{\epsilon^2}Q(t)_{i,i} =
  -\frac{1}{2}\dot{\mathbf{e}}(t)\dot{\mathbf{e}}(t)\,, \\
&
\frac{1}{\epsilon^2}Q(t)_{i0}Q(t)_{i0} =
\dot{\mathbf{e}}(t)\dot{\mathbf{e}}(t)\,\,.
\end{split}
\end{equation}
Let us demonstrate eq.~\eqref{25} .\\
\newline
$\frac{1}{\epsilon^2}\underline{Q(t)_{00}}$ \\
\newline
Using eqs.~\eqref{23},~\eqref{20} and ${\Sigma}_{0,a}=\delta_{0,a}$ we obtain
\begin{equation}
\label{26}
\frac{1}{\epsilon^2}{Q(t)_{00}}=\frac{1}{\epsilon^2}(V(t+\epsilon)-1)_{00}=
 \frac{1}{\epsilon^2}(\Omega(t)^T\Omega(t+\epsilon)-1)_{00}\,\,.
\end{equation}
Expanding in $\epsilon$ gives
 \begin{equation}
  \label{27}
\half [\Omega(t)^T \ddot{\Omega}(t)]_{00}=
 -\half [\dot{\Omega}(t)^T \dot{\Omega}(t)]_{00}=
 -\half  \dot{\mathbf{e}}(t)\dot{\mathbf{e}}(t)\,.
\end{equation}
Here we used 
\begin{equation}
\label{271}
2 \dot{\Omega}(t)^T\dot{\Omega}(t)+ \ddot{\Omega}(t)^T\Omega(t) +
\Omega(t)^T\ddot{\Omega}(t) =0\,.
\end{equation}
Turn to the next case: \\
\newline
$\frac{1}{\epsilon^2}\underline {Q(t)_{i0}}\underline {Q(t)_{i0}} $ \\
\newline
Consider first
\begin{equation}
 \label{28}
 \frac{1}{\epsilon}Q(t)_{a0}=\frac{1}{\epsilon}[{\Sigma}(t) \Omega(t)^T 
 \Omega(t+\epsilon)-{\Sigma}(t)]_{a0} = [{\Sigma}(t) \Omega^T(t)]_{ab} 
 \,\dot{e}(t)_{b} \,,
\end{equation}
where we used that ${\Sigma}(t)_{a0}=\delta_{a0}$.
It follows then
\begin{equation}
 \label{29}
 \frac{1}{\epsilon}Q(t)_{i0} \frac{1}{\epsilon}Q(t)_{i0}=
 \frac{1}{\epsilon}Q(t)_{a0} \frac{1}{\epsilon}Q(t)_{a0}= 
 \dot{\mathbf{e}}(t)\dot{\mathbf{e}}(t) \,.
\end{equation}
Here we used that $\frac{1}{\epsilon}Q(t)_{00}$ is $O(\epsilon)$. \\
\newline
$\frac{1}{\epsilon^2}\underline {Q(t)_{ii}}$ \\
\newline
eq.~\eqref{23} implies
\begin{equation}
\label{30}
\frac{1}{\epsilon^2}Q(t)_{ii}=
-\frac{1}{\epsilon^2} \frac{ V(t+\epsilon)_{i0} V(t+\epsilon)_{i0} } 
{1+V(t+\epsilon)_{00}}\,.  
\end{equation}
Here 
$V(t+\epsilon)_{i0}= Q(t)_{i0}$
and the denominator is $2+O(\epsilon^2)$, we obtain from eq.~\eqref{29} the
result eq.~\eqref{25}.

As discussed before, the fast modes are carried by the
${\mathbf{R}}=[(1-{\mathbf {\Pi}^2})^\half,{\mathbf {\Pi}}]$ field, where the 
$k=(k_0,\mathbf{k}=\mathbf{0})$ modes are missing. The first term on the r.h.s. of
eq.~\eqref{19} gives the fast $\Pi-\Pi$ interactions, while the second term
describes the fast-low interactions. We get for the leading action 
$A^{(2)}_\mathrm{eff}$ in eq.~\eqref{5}:
\be
\label{31}
\begin{split}
&A^{(2)}_\mathrm{eff}(\Omega {\Sigma}^T {\mathbf{R}}) = \\
&\int_x  \frac{F^2}{2}
\Big\{\partial_\mu {\mathbf{R}}(x)\, \partial_\mu
{\mathbf{R}}(x)   \\
& -\frac{2}{\epsilon^2}Q(t)_{00}[1-{\mathbf {\Pi}^2}(x)]
-\frac{2}{\epsilon^2}Q(t)_{ij}\Pi(x)_i\, \Pi(x)_j  \\
&  -\frac{2}{\epsilon}Q(t)_{i0}
 \big[\partial_t(1-{\mathbf {\Pi}^{2}}(x))^\half \,\Pi(x)_i
-\partial_t \Pi(x)_i\,(1-{\mathbf {\Pi}^2}(x))^\half \big] \Big\}  \, \,.
\end{split}
\ee
The terms above  give the action  $A^{(2)}_\mathrm{eff}$ up to NNL
order. We shall expand in the $\Pi$-fields in perturbation theory. The last
term in eq.~\eqref{31}, which is odd in $\Pi$, 
enters only in the NNL order of this expansion.

\subsection{Counting rules}
 The small expansion parameter in the $\delta$-regime is
$1/F^2 L^2_s = O(\delta^2)$. The expansion of the rotator action has the form
\be
\label{32}
\int_t \frac{F^2}{2} V_s \dot{\mathbf{e}}(t)\dot{\mathbf{e}}(t)
(1+\sim\frac{1}{F^2 L^2_s}+ \sim \frac{1}{(F^2 L^2_s)^2} \dots)
\ee
in the leading L, next-to-leading NL, NNL,$\dots$ order.

The finite part of the pairings of the fast modes are $< \Pi\,\Pi >$ $\sim 1/(F^2
  L^2_s)$ and  $< \partial_{\mu}\Pi\partial_{\nu} \Pi >$ $\sim
  \delta_{\mu,\nu}/(F^2 L^4_s)$. 

In the expansion there are terms also with quadratic and higher powers of the
slow mode $\dot{\mathbf{e}}$. The following consideration shows that an
additional $\sim \dot{\mathbf{e}} \dot{\mathbf{e}}$ term in the bracket of 
eq.~\eqref{32} is $O(\delta^6)$, i.e. beyond our NNL calculation.

The argument is as follows. The leading Lagrangian is 
$\sim F^2 L^3_s \dot{\mathbf{e}} \dot{\mathbf{e}}$
which gives for the conjugate momentum ${\mathbf{L}} \sim F^2
L^3_s\, \dot{\mathbf{e}}$. It follows then $\dot{\mathbf{e}} \dot{\mathbf{e}}\sim
 {\mathbf{L}^2} (F^2 L^3_s)^{-2}$. Since ${\mathbf{L}^2}\sim O(1)$, we obtain
\be
\label{33}
\dot{\mathbf{e}}(t)\dot{\mathbf{e}}(t)L^2_s \sim (\frac{1}{F^2
  L^2_s})^2=O(\delta^4)\,, 
\ee
giving $\dot{\mathbf{e}} \dot{\mathbf{e}}\sim O(\delta^6)$ as stated
above. In the NNNL order, which is beyond our calculation, such corrections
are expected to enter. Consider an example. As we shall see later, in the
Boltzmann factor enters (among others)
\begin{equation}
\label{331}
\begin{split}
&\exp \Big[\int_x \frac{F^2}{2} \dot{\mathbf{e}}(t)\dot{\mathbf{e}}(t)
{\mathbf{\Pi}}(x){\mathbf{\Pi}}(x)\Big]=
1+\int_x \frac{F^2}{2} \dot{\mathbf{e}}(t)\dot{\mathbf{e}}(t)
{\mathbf{\Pi}}(x){\mathbf{\Pi}}(x) \\
&+\frac{1}{2}\int_x \frac{F^2}{2} \dot{\mathbf{e}}(t)\dot{\mathbf{e}}(t)
{\mathbf{\Pi}}(x){\mathbf{\Pi}}(x)
\int_{x'} \frac{F^2}{2} \dot{\mathbf{e}}(t')\dot{\mathbf{e}}(t')
{\mathbf{\Pi}}(x'){\mathbf{\Pi}}(x')+ \dots \,.
\end{split}
\end{equation}
We pair out the fast modes on the r.h.s. of eq.~\eqref{331}. Pairing $\Pi(x)$ with
$\Pi(x)$ and $\Pi(x')$ with $\Pi(x')$ in the second line is just needed for the
exponentialization. Pairing  $\Pi(x)$ with $\Pi(x')$ gives, however
\begin{equation}
\label{332}
\begin{split}
&\sim \int_t \int_{t'} [\dot{\mathbf{e}}(t) \dot{\mathbf{e}}(t)]\,
[\dot{\mathbf{e}}(t')\dot{\mathbf{e}}(t')] \int_\mathbf{x} \int_\mathbf{x'}
D^*(x-x')^2 = \\
&\int_t \int_{t''} F^2 L_s^3\, [\dot{\mathbf{e}}(t)\dot{\mathbf{e}}(t)]\,
[\dot{\mathbf{e}}(t-t'')\dot{\mathbf{e}}(t-t'')] \int_{x''}
D^*(x'')^2 \frac{1}{F^2}\,.
\end{split}
\end{equation}
In $D^*(x'')$ the ${\mathbf{k}=0}$ is missing, the smallest ${\mathbf{k}}$ is
$\sim 1/L_s$. Therefore,  $D^*(x'')$ has an exponential cut on the
level $t-t''\sim L_s$. The time distance $L_s$ is small for the rotator,
therefore $\dot{\mathbf{e}}(t-t'') \sim \dot{\mathbf{e}}(t)$. The integral
over $D^*$, after renormalization, is $O(1)$ giving
\begin{equation}
\label{333}
\sim \int_t  F^2 L_s^3 [\dot{\mathbf{e}}(t) \dot{\mathbf{e}}(t)]
\frac{1}{F^2} [\dot{\mathbf{e}}(t) \dot{\mathbf{e}}(t)] = O(\delta^6)\,.
\end{equation}

 \subsection{The action {$A^{(4)}_\mathrm{eff}(\Omega {\Sigma}^T {\mathbf{R}})$}}
In the term with $~l_1$ in eq.~\eqref{5} we can use the result in eq.~\eqref{19}
\begin{equation}
\label{34}
\begin{split}
&\partial_\mu \mathbf{S}(x) \partial_\mu \mathbf{S}(x)\,
\partial_\nu \mathbf{S}(x)\partial_\nu \mathbf{S}(x)=
 \partial_\mu \mathbf{R}(x)\partial_\mu \mathbf{R}(x)\,
\partial_\nu \mathbf{R}(x)\partial_\nu \mathbf{R}(x)\,-\\
&2\,\partial_\mu \mathbf{R}(x)\partial_\mu \mathbf{R}(x)\,
\frac{2}{\epsilon^2}
Q(t)\mathbf{R}(t+\epsilon,\mathbf{x})\,\mathbf{R}(t,\mathbf{x})\,+\\
& \frac{2}{\epsilon^2}
Q(t)\mathbf{R}(t+\epsilon,\mathbf{x})\mathbf{R}(t,\mathbf{x})\,
 \frac{2}{\epsilon^2} 
Q(t)\mathbf{R}(t+\epsilon,\mathbf{x})
\mathbf{R}(t,\mathbf{x})_{\epsilon \rightarrow 0} \,.
\end{split}
\end{equation}
The leading term of the first part on the r.h.s. in eq.~\eqref{34} is
$\partial_\mu \mathbf{\Pi} \partial_\mu \mathbf{\Pi}\,
\partial_\nu \mathbf{\Pi} \partial_\nu \mathbf{\Pi}$. This term can influence
the low mode spectrum only if it is multiplied with both low and
fast modes like $~\dot{\mathbf{e}}\dot{\mathbf{e}}\mathbf{\Pi}\,\mathbf{\Pi}$.
This combination is is far beyond our calculation. Up to NNL order we can write 
for the $~l_1$ part
\begin{equation}
\label{35}
\begin{split}
\partial_\mu \mathbf{S}(x) \partial_\mu \mathbf{S}(x)\,
\partial_\nu \mathbf{S}(x) \partial_\nu \mathbf{S}(x)& =
 -\frac{4}{\epsilon^2} Q(t)_{00}\,\, \partial_\mu \mathbf{\Pi}(x) 
\partial_\mu \mathbf{\Pi}(x)\,\\
&  +\frac{4}{\epsilon^2} Q(t)_{i0}\,\, Q(t)_{j0} \, 
  \partial_0 \Pi(x)_i \,\partial_0 \Pi(x)_j \,.
\end{split}
\end{equation}
The term with $~l_2$ in eq.~\eqref{5} requests somewhat more, but trivial work. It
has the form
\begin{equation}
\label{36}
\begin{split}
&\partial_\mu \mathbf{S}(x) \partial_\nu \mathbf{S}(x)\,
\partial_\mu \mathbf{S}(x)\partial_\nu \mathbf{S}(x) =
 -\frac{4}{\epsilon^2}Q(t)_{00}\,\, \partial_0 \mathbf{\Pi}(x) \,\partial_0
\mathbf{\Pi}(x) \\
&+\frac{4}{\epsilon^2} Q(t)_{i0}\,Q(t)_{j0}\Big\{ \partial_0 \Pi(x)_i \,\partial_0
\Pi(x)_j +\half \sum^3_{\alpha=1}\partial_{\alpha}\Pi(x)_i\,
\partial_{\alpha}\Pi(x)_j \Big\}\,.
\end{split}
\end{equation}
The part of the action $A^{(4)}_\mathrm{eff}$ which is relevant up to NNL order
has the form
\begin{equation}
\label{35}
\begin{split}
&A^{(4)}_\mathrm{eff}(\Omega {\Sigma}^T {\mathbf{R}})=\\
&\int_x \Big \{
{l_1} \Big [ \frac{4}{\epsilon^2}Q(t)_{00}\,\, \partial_0 \mathbf{\Pi}(x)
\,\partial_0 \mathbf{\Pi}(x)
-\frac{4}{\epsilon^2} Q(t)_{i0}\,Q(t)_{j0} \partial_0 \Pi(x)_i \,\partial_0
\Pi(x)_j \Big ] \\
&+{l_2} \Big [
-\frac{4}{\epsilon^2} Q(t)_{i0}\,Q(t)_{j0} \Big( \partial_0 \Pi(x)_i
\,\partial_0 \Pi(x)_j +\half \sum^3_{\alpha=1}\partial_{\alpha}\Pi(x)_i
\partial_{\alpha}\Pi(x)_j \Big ) \\
&\frac{4}{\epsilon^2}Q(t)_{00}\,\, \partial_0 \mathbf{\Pi}(x) \,\partial_0
\mathbf{\Pi}(x) \Big ]
\Big \}_{\epsilon \rightarrow 0}\,\,.
\end{split}
\end{equation}

\section{The path integral up to NNL order}
The relevant part of the action is the sum of the terms on the r.h.s. of the
equations~\eqref{31} and ~\eqref{35}. This action goes in the path integral as
${\rm exp}(-A^{(2)}_\mathrm{eff}-A^{(4)}_\mathrm{eff})$, where the action depends
on the slow ${\mathbf{e(t)}}$ and the fast $\mathbf{\Pi}(x)$ degrees of freedom. 
We consider the partition function eq.~\eqref{16} in dimensional
regularization (DR,${\overline{MS}}$). Among other conveniences, this
simplifies the measure. The Boltzmann factor  
${\rm exp}(-A^{(2)}_\mathrm{eff}-A^{(4)}_\mathrm{eff})$ is expanded in the
$\mathbf{\Pi}$ fields: only the leading parts remain in the exponent, the rest
is a Taylor expansion in powers of $\mathbf{\Pi}$. 

The path integral has the following form up to NNL order\footnote{To make sure
  the notation: the
  brackets $\Big\{ \dots \Big \}$ are multiplied with each other.}:
\begin{equation}
\label{36}
\begin{split}
&Z=\\
&\prod_t \int d{\mathbf{e}}(t) \prod_{\mathbf{x}} 
\int d{\mathbf{\Pi}}(t,{\mathbf{x}}) \prod_{i=1}^{N-1}\delta
(\frac{1}{V_s}\int_{\mathbf{y}}\Pi_i(t,{\mathbf{y}})) \\
&\exp \Big[{-\int_t \frac{F^2}{2}V_s \dot{\mathbf{e}}(t)
\dot{\mathbf{e}}(t)}\Big]\,\,
\exp \Big[{-\int_x \frac{F^2}{2} \partial_{\mu} {\mathbf{\Pi}}(x) \,\partial_{\mu}
{\mathbf{\Pi}(x)}}\Big] \\
(1) \qquad &\Big \{ 1+\int_x \frac{F^2}{2} \dot{\mathbf{e}}(t)\dot{\mathbf{e}}(t)
{\mathbf{\Pi}^2}(x) + \dots
\Big \}   \\
(2) \qquad &\Big \{ 1+\int_x \frac{F^2}{2}\frac{2}
{\epsilon^2}Q(t)_{ij}\Pi(x)_i \Pi(x)_j + \dots
 \Big \} \\
(3) \qquad &\Big \{ 1+\frac{1}{2!} \Big [ 
\int_x F^2 \frac{1}{\epsilon}Q(t)_{i0} \Big(\half {\mathbf{\Pi}^2}(x)
\partial_0 \Pi(x)_i - {\mathbf{\Pi}}(x)\partial_0{\mathbf{\Pi}}(x)\Pi(x)_i
\Big ) \\
& \int_y F^2 \frac{1}{\epsilon}Q(y_0)_{j0} \Big(\half {\mathbf{\Pi}^2}(y)
\partial_0 \Pi(y)_j - {\mathbf{\Pi}}(y)\partial_0{\mathbf{\Pi}}(y)\Pi(y)_j
\Big ) \Big ] + \dots 
\Big \}  \\
(4) \qquad &\Big \{ 1-\int_x \frac{F^2}{2}(\mathbf{\Pi}(x)
\partial_{\mu}{\mathbf{\Pi}}(x))
({\mathbf{\Pi}}(x)\partial_{\mu}{\mathbf{\Pi}}(x)) + \dots
\Big \} \\
(5) \qquad &\Big \{ 1+4\,l_1 \int_x \half \dot{\mathbf{e}}(t)\dot{\mathbf{e}}(t)
\partial_{\mu} {\mathbf{\Pi}}(x) \,\partial_{\mu} {\mathbf{\Pi}}(x) \\
& + \frac{1}{\epsilon^2}Q(t)_{i0}Q(t)_{j0}\, 
\partial_0 \Pi(x)_i \partial_0 \Pi(x)_j + \dots
\Big \} \\
(6) \qquad &\Big \{ 1+4\,l_2 \int_x \half \dot{\mathbf{e}}(t)\dot{\mathbf{e}}(t)
\partial_0 {\mathbf{\Pi}}(x) \,\partial_0 {\mathbf{\Pi}}(x)
+\frac{1}{\epsilon^2}Q(t)_{i0}Q(t)_{j0} \\
&\Big (\partial_0 \Pi(x)_i
\,\partial_0 \Pi(x)_j +\half \sum^3_{\alpha=1}\partial_{\alpha}\Pi(x)_i
\partial_{\alpha}\Pi(x)_j \Big ) + \dots
\Big \}\,.
\end{split}
\end{equation}

The integration variables in this path integral are the slow modes
${\mathbf{e}}(t)$ and the fast $\mathbf{\Pi}(x)$ modes with the constraint
$1/V_s \int_{\mathbf{x}}  \Pi(t,{\mathbf{x}})_i=0$, i.e. in Fourier space the 
$(k_0,{\mathbf{k}}=0)$-modes are missing. The corresponding
pairings are
\be
\label{37}
< \Pi(x)_i \Pi(0)_j > = \frac{\delta_{i,j}}{F^2} D^{\ast}(x)\,,
< \partial_0 \partial_0 \Pi(x)_i \Pi(0)_j > = \frac{\delta_{i,j}}{F^2}\, 
\partial_0 \partial_0 D^{\ast}(x)     \,.
\ee

In eq.~\eqref{36} the six lines numbered as $(1) \dots (6)$ correspond to the
expansion of the action in the exponent. We gave explicitly the terms only which are
needed up to NNL. 
We integrate out in eq.~\eqref{36} the $\Pi$-fields and obtain
a path integral in quantum mechanics for a rotator. The first two
contributions below are NL\footnote{The NL result is in agreement with that 
in~\cite{PHFN}.}, the other five are NNL:
$\newline$\\
$(1)$\\
$ \quad \Big \{1+\int_t \half F^2 V_s \dot{\mathbf{e}}(t)
\dot{\mathbf{e}}(t) \frac{N-1}{F^2} D^{\ast}(0) \Big \}$ \\
$\newline$
$(2)$\\
$ \quad \Big \{1-\int_t \half F^2 V_s \dot{\mathbf{e}}(t)\dot{\mathbf{e}}(t)
\frac{1}{F^2} D^{\ast}(0) \Big \}$\\
\newline
$(1)-(4)$ crossing \\
$ \quad \Big \{1-\int_t \half F^2 V_s \dot{\mathbf{e}}(t)\dot{\mathbf{e}}(t)
\frac{N-1}{F^4} D^{\ast}(0)D^{\ast}(0)\Big \}$\\
$\newline$
$(2)-(4)$ crossing \\
$ \quad \Big \{1+\int_t \half F^2 V_s \dot{\mathbf{e}}(t)\dot{\mathbf{e}}(t)
\frac{1}{F^4} D^{\ast}(0)D^{\ast}(0)\Big \}$\\
$\newline$
$(3)$\\
$ \quad \Big \{1-\int_t \half F^2 V_s \dot{\mathbf{e}}(t)\dot{\mathbf{e}}(t)
\frac{2N-4}{F^4}\int_z
\partial_0 \partial_0 D^{\ast}(z) D^{\ast}(z)D^{\ast}(z) \Big \}$\\
$\newline$
$(5)$\\
$ \quad \Big \{1-4\,l_1\int_t \half F^2 V_s
\dot{\mathbf{e}}(t)\dot{\mathbf{e}}(t) 
\frac{2}{F^4}\, \partial_0 \partial_0 D^{\ast}(0) \Big \}$\\ 
$\newline$
$(6)$\\
$ \quad \Big \{1-4\,l_2\int_t \half F^2 V_s
\dot{\mathbf{e}}(t)\dot{\mathbf{e}}(t) 
\frac{N}{F^4}\,  \partial_0 \partial_0 D^{\ast}(0) \Big \}$\\ 

Bringing these contributions in the exponent,
we get a standard rotator like in 
eq.~\eqref{7}. Only the inertia $\Theta$ has corrections:
\begin{equation}
\label{38}
\begin{split}
\Theta = &
F^2V_s \Big \{
 1-\frac{N-2}{F^2} D^{\ast}(0)+ \frac{N-2}{F^4}
 D^{\ast}(0) D^{\ast}(0)\\
&+ 2 \frac{N-2}{F^4}\int_x \partial_0 \partial_0 D^{\ast}(x)
D^{\ast}(x) D^{\ast}(x)  +\frac{1}{F^4}(8l_1 + 4Nl_2)
\partial_0 \partial_0 D^{\ast}(0)\,.
\Big \}
\end{split}
\end{equation}
In the following we consider $N=4$, corresponding to two-flavor QCD.
Since $D^{\ast}(0)$ and $\partial_0 \partial_0 D^{\ast}(0)$ are known, the
only remaining task is to determine the integral in eq.~\eqref{38}. This
integral is UV-divergent. The singularity should be canceled by the divergent
part of the low energy constants $l_1$ and $l_2$. The singularities of these
constants are known in dimensional regularization~\cite{JGHL2}.

\section{Green's functions}
We are in $d$ dimension, $d=4+\epsilon$. The physical space-time in Euclidean
space is  $L_s \times L_s \times L_s \times (L_t \rightarrow \infty)$ and the
corresponding Green's function is $D(x;d)$. Subtracting the ultraviolet and
infrared (notation: 'bar' and 'star', respectively) divergences we obtain a
finite Green's function
\be
\label{55}
{\bar{D}}^*(0) = D(0;d)-{(2\pi)^d}\int dk^d \frac{1}{k^2+m^2} -
  \Big( \frac{1}{L_s}  \Big)^3  (2\pi)^{d-3} \int dk^{d-3} \,
\frac{1}{k^2+m^2}\,.
\ee
We consider the chiral limit $m\rightarrow 0$. The UV subtraction (the second
term on the r.h.s. above) is zero in dimensional regularization . Therefore,
${\bar{D}}^*(0)=D^*(0)$ and $\partial_0\partial_0{\bar{D}}^*(0)=
\partial_0\partial_0 D^*(0)$. For the latter, in the framework of DR, 
the IR subtraction is zero also:  
$\partial_0\partial_0{\bar{D}}^*(0)=\partial_0\partial_0 D(0)$.

For numerical purposes the following representation is useful\footnote{See,
  for example, in\cite{PGHL,PHHL}.}
\be
\label{56}
{\bar{D}}^*(x)
= \frac{1}{L_s^2} \int_0^\infty dw \frac{1}{4\pi} e^{-\pi w y_0^2}\Big \{ 
\Big [\Pi^{3}_{i=1}
S(w,y_i) -  e^{-\pi w \mathbf{y}^2}\Big ] -w^{-\frac{3}{2}} \Big \}\,,
\ee
where  $y=x/L_s$ and
\begin{eqnarray}
\label{57}
S(w,z)&=& w^{-\half} \sum^{\infty}_{n=-\infty} e^{-\pi\frac{1}{w}n^2} 
cos(2\pi n z)\,,\qquad 0 < w <1 \,,\\
S(w,z)&=&  \sum^{\infty}_{n=-\infty} e^{-\pi w(n+z)^2} \,,\qquad 
1 < w <\infty \,. 
\end{eqnarray}

\section{The integral $\int_{\tau} dx\, \partial_0 \partial_0 D^{\ast}(x)
D^{\ast}(x) D^{\ast}(x)$ }
We calculate this integral using dimensional regularization (DR) in the chiral
limit.The region of integration is $L_s \times L_s \times L_s \times (L_t
\rightarrow \infty)$. We write
\be
\label{39}
D^{\ast}(x) =\Delta(x) + {\bar D^{\ast}}(x)\,\,, 
\ee 
where $\Delta(x)$ is the infinite volume propagator
\be
\label{40}
\Delta(x)=\frac{1}{4\pi^2 r^2}\,,\,\,\,\, 
\partial_0 \partial_0 \Delta(x)=\frac{1}{4\pi^2} (4x_0^2-r^2)\frac{2}{r^6}\,\,.
\ee 
The Green's functions ${\bar D^{\ast}}(x)$ and $\partial_0\partial_0{\bar D^{\ast}}(x)$
are free of IR and UV singularities. 

The integral falls then into six terms. The following four terms (1,2,3,6)
are finite in DR :  
\be
\label{41}
\begin{split}
\int_{\tau} dx\,\Big \{ & 
\partial_0 \partial_0 {\bar D^{\ast}}(x) 
{\bar D^{\ast}}(x) {\bar D^{\ast}}(x) + 
\partial_0 \partial_0 \Delta(x) {\bar D^{\ast}}(x) {\bar D^{\ast}}(x) +\\
& 2\,\Delta(x) \partial_0 \partial_0 {\bar D^{\ast}}(x) {\bar D^{\ast}}(x) +
\partial_0 \partial_0 \Delta(x) \Delta(x) \Delta(x) 
\Big \}\,\,,
\end{split}
\ee
while the terms below (4,5) are UV singular:
\be
\label{42}
\int_{\tau} dx\,\Big \{ 
\Delta(x)  \Delta(x) \partial_0 \partial_0 {\bar D^{\ast}}(x) +
2\,\partial_0 \partial_0 \Delta(x) \Delta(x) {\bar D^{\ast}}(x)
\Big \}\,\,.
\ee

It is useful to divide the integration region $\tau$ into a 'cube' and
a 'left-right' region, where 'cube'= $(-L_s/2,L_s/2)^4$ and
 'right'=$(-L_s/2,L_s/2)^3 \,,(L_s,L_t)$ while 
 'left'=$(-L_s/2,L_s/2)^3 \,,(-L_s,-L_t)$. The left-right region is 
UV-safe\footnote{The procedure applied here is similar to that used
  in~\cite{PGHL,PHHL}.}.

Consider the integrals in eq.~\eqref{41}. The first and third integrals are
finite and can be integrated as they are. For the second integral we write
\be
\label{43}
\begin{split}
&\int_{\tau} dx\, \partial_0 \partial_0 \Delta(x) {\bar D^{\ast}}(x) 
{\bar D^{\ast}}(x)=\\
&{\bar D^{\ast}}(0)^2 \int_{\tau} dx\, \partial_0 \partial_0 \Delta(x) +
\int_{\tau} dx\,\partial_0 \partial_0 \Delta(x) \Big ( 
{\bar D^{\ast}}(x)^2-{\bar D^{\ast}}(0)^2
\Big )
\end{split}
\ee
The first integral on the r.h.s. of  eq.~\eqref{43} is zero in the cube due to
$90^\circ$ rotation symmetry. The rest in eq.~\eqref{43} can be integrated as it is.
The fourth integral in  eq.~\eqref{41} is also zero over the cube. The rest is
finite. Let us list the four integral values (1,2,3,6) with $L_s^{-4}$ suppressed:
\be
\label{431}
-1.228902057d-2\,,\quad -1.730322906d-2\,,\quad 2.461536207d-2\,,\quad
4.1588904d-4\,. 
\end{equation} 
The final sum of the four integrals in eq.~\eqref{41} is\,\,
\be
\label{44}
-\frac{1}{L_s^4} 0.00456099852 \,.
\ee

Turn now to the singular integrals in eq.~\eqref{42}. Write the integral (4) as
\be
\label{45}
\begin{split}
\partial_0 \partial_0 {\bar D^{\ast}}(0)
\Big \{&
\int_R \Delta(x) \Delta(x) - \int_{R\setminus S} \Delta(x) \Delta(x) \\
& +\int_{\tau \setminus S} \Delta(x) \Delta(x)
\Big \} +
\int_{\tau} \Delta(x) \Delta(x) \Big (
\partial_0 \partial_0 {\bar D^{\ast}}(x) - \partial_0 \partial_0 {\bar
  D^{\ast}}(0)
\Big )\,,
\end{split}
\ee
where $R \setminus S$ is the full space-time with a sphere cut around $x=0$
and similarly for $\tau \setminus S$. The radius
of the sphere is less then, or equal to $L_s$. 
Only the first integral in eq.~\eqref{45} is singular with DR, the rest is
finite. The result is for (4)
\be
\label{46}
-\frac{1}{L_s^4}\Big \{ d0d0{\bar G}^{\ast} \frac{1}{8\pi^2} 
\Big [\frac{1}{d-4}+
\ln(\frac{1}{L_s}) \Big ] +0.009856387107
\Big \}\,.
\ee
 
In the second integrand (5) in eq.~\eqref{42} we follow the steps applied above:
expand ${\bar D^{\ast}}(x)$ around $x=0$ until the rest gives a finite
integral. We write
\be
\label{47}
\begin{split}
{\bar D^{\ast}}(x)=& {\bar D^{\ast}}(0) + \frac{1}{2V_s}|x_0| +
\half \sum_{\mu\,,\nu}\Big ( \partial_{\mu} \partial_{\nu}{\bar D^{\ast}}(0) x_{\mu}x_{\nu}\Big ) \\
& + \Big \{
{\bar D^{\ast}}(x) -{\bar D^{\ast}}(0) - \frac{1}{2V_s}|x_0| -
\half \sum_{\mu\,,\nu}\Big ( \partial_{\mu} \partial_{\nu}{\bar D^{\ast}}(0) x_{\mu}x_{\nu}\Big )
\Big \}\,.
\end{split}
\ee
Only the third term on the r.h.s. of eq.~\eqref{47} is singular in DR.
The result is
\be
\label{49}
-\frac{1}{L_s^4}\Big \{ d0d0{\bar G}^{\ast} \frac{1}{8\pi^2}\frac{2}{3} \Big
[\frac{1}{d-4}+ 
\ln(\frac{1}{L_s}) \Big ] +0.015074639535
\Big \}\,.
\ee
Collecting the results from eqs.~\eqref{44},~\eqref{46},~\eqref{49} we obtain
\be
\label{50}
\begin{split}
&\int_{\tau} dx\, \partial_0 \partial_0 D^{\ast}(x)
D^{\ast}(x) D^{\ast}(x)= \\
&-\frac{1}{L_s^4}\Big \{ d0d0{\bar G}^{\ast} \frac{1}{8\pi^2}\frac{5}{3} \Big
[\frac{1}{d-4}+ 
\ln(\frac{1}{L_s}) \Big ] + 0.029492025146 \,.
\Big \}
\end{split}
\ee

\section{The final result on the inertia $\Theta$}
The singular part of the bare low-energy constants $l_i,i=1,2$ are 
known in dimensional regularization~\cite{JGHL2}. For the finite part 
we follow a generally accepted convention~\cite{CGL}.
The bare low-energy constants $l_i$ are written as
\be
\label{51}
l_i=\gamma_i\, \lambda + l_i^r
\ee
where $\gamma_1=\frac{1}{3}$, $\gamma_2=\frac{2}{3}$ and
\be
\label{52}
\lambda=\frac{1}{16\pi^2}
\Big \{\frac{1}{d-4}-\half\, \Big(\ln(4\pi)-C+1-\ln(\mu^2)\,\Big)\Big\}\,\,.
\ee
Write the renormalized coupling constants as
\be
\label{53}
l_i^r = \frac{\gamma_i}{32\pi^2}\ln(\frac{\Lambda^2_i}{\mu^2})\,\,,
\ee
which gives
\be
\label{54}
l_i=\frac{\gamma_i}{16\pi^2}\frac{1}{d-4} -\frac{\gamma_i}{32\pi^2}
\Big( \ln(4\pi)-C +1-\ln(\Lambda_i^2)\Big)\,. 
\ee
The combination which enters in eq.~\eqref{38} is
\be
\label{541}
\begin{split}
\frac{d0d0{\bar G}^*}{(F^2L_s^2)^2}&(8l_1+16l_2)=
\frac{d0d0{\bar G}^*}{(F^2L_s^2)^2}
\Big\{\frac{5}{6\pi^2}\frac{1}{d-4}\\
& -\frac{5}{12\pi^2}\Big(\ln(4\pi)-C +1\Big)+
\frac{1}{3\pi^2}\Big(\frac{1}{4}\ln(\Lambda_1^2)+\ln(\Lambda_2^2\Big)
\Big\}\,.
\end{split}
\ee

Multiplying the result in eq.~\eqref{50} by a factor of 4 and adding to
eq.~\eqref{541} we find that the singularities in eq.~\eqref{38} 
cancel and the result in eq.~\eqref{2} is obtained.

\noindent{\bf Acknowledgments} \\
The author is indebted for discussions with G.~Colangelo, 
H.~Leutwyler, F.~Niedermayer, M.~Weingart and Ch.~Weyermann.
This work is supported in part by the Schweizerischer Nationalfonds.
The author acknowledges support by DFG project SFB/TR-55. 
The "Albert Einstein Center for Fundamental Physics" at Bern University is
supported by the "Innovations-und Kooperationsprojekt C-13"
of the Schweizerischer  Nationalfonds.




\newpage


 \begin{thebibliography}{99}

 \bibitem{Le1}
 H.~Leutwyler, Phys.~Lett.~{\bf B189}, 197 (1987).

 \bibitem{SW}
 S.~Weinberg, Physica~{\bf A96}, 327 (1979).

 \bibitem{JGHL1}
 J.~Gasser and H.~Leutwyler, Phys.~Lett.~{\bf B125}, 321,325 (1983).

 \bibitem{JGHL2}
 J.~Gasser and H.~Leutwyler, Ann.~Phys.~{\bf 158}, 142 (1984).

 \bibitem{FiPr}
 M.~E.~Fisher and V.~Privman, Phys.~Rev.~{\bf B32}, 447 (1985).

 \bibitem{BEZJ}
 E.~Brezen and J.~Zinn-Justin, Nucl.~Phys.~{\bf B257}, 867 (1985).
 
 \bibitem{PHFN}
 P.~Hasenfratz and F.~Niedermayer, Z.~Phys.~{\bf B92}, 91 (1993).

 \bibitem{FNCW}
 F.~Niedermayer and Ch.~Weiermann, in progress.

 \bibitem{MW}
 M.~Weingart, in progress.

 \bibitem{JGHL3}
 J.~Gasser and H.~Leutwyler, Phys.~Lett.~{\bf B184}, 83 (1987).

 \bibitem{JGHL4}
 J.~Gasser and H.~Leutwyler, Phys.~Lett.~{\bf B188}, 477 (1987).

 \bibitem{JGHL5}
 J.~Gasser and H.~Leutwyler, Nucl.~Phys.~{\bf B307}, 763 (1988).

 \bibitem{PHHL}
  P.~Hasenfratz and H.~Leutwyler, Nucl.~Phys.~{B343}, 241 (1990).

 \bibitem{PGHL}
 P.~Gerber and H.~Leutwyler, Nucl.~Phys.~{\bf B321}, 387 (1989).

 \bibitem{CGL}
 G.~Colangelo, J.~Gasser and H.~Leutwyler, Nucl.~Phys.~{\bf B603},125 (2001).











 \end{thebibliography}
\end{document}